\documentstyle[prl,aps,multicol,epsf]{revtex}
\renewcommand{\narrowtext}{\begin{multicols}{2} \global\columnwidth20.5pc}
\renewcommand{\widetext}{\end{multicols} \global\columnwidth42.5pc}

\begin{document}
\newcommand{\be}{\begin{equation}}
\newcommand{\ee}{\end{equation}}
\newcommand{\ba}{\begin{eqnarray}}
\newcommand{\ea}{\end{eqnarray}}
\newcommand{\abar}{\overline{a}}
\newcommand{\bbar}{\overline{b}}
\newcommand{\cbar}{\overline{c}}
\newcommand{\fbar}{\overline{f}}
\newcommand{\ubar}{\overline{u}}
\newcommand{\vbar}{\overline{v}}
\newcommand{\Gbar}{\overline{\Gamma}}
\newcommand{\gbar}{\overline{\gamma}}
\newcommand{\Dij}{\Delta_{ij}}
\newcommand{\Dbarij}{\overline{\Delta}_{ij}}
\newcommand{\p}{\partial_0}
\newcommand{\aavg}{\overline{\rho}_a}
\newcommand{\bavg}{\overline{\rho}_b}
\newcommand{\cavg}{\overline{\rho}_c}
\newcommand{\phibar}{\overline{\phi}}
\newcommand{\psibari}{\overline{\psi}_{i\sigma}}
\newcommand{\psibar}{\overline{\psi}}
\newcommand{\psii}{\psi_{i\sigma}}
\newcommand{\psij}{\psi_{j\sigma}}
\newcommand{\0}{\partial_0}
\newcommand{\ij}{\langle ij \rangle}
\newcommand{\Kij}{K_{ij}}
\newcommand{\Kbarij}{\overline{K}_{ij}}
\newcommand{\Ubarij}{\overline{U}_{ij}}
\newcommand{\Pbarij}{\overline{P}_{ij}}
\newcommand{\Hbarij}{\overline{H}_{ij}}
\newcommand{\Hij}{H_{ij}}
\newcommand{\Pij}{P_{ij}}
\newcommand{\Si}{{\bf S}_i}
\newcommand{\Sj}{{\bf S}_j}
\title{The Structure of a Vortex in the t-J Model}
\author{Jung Hoon Han and Dung-Hai Lee}
\address{Department of Physics, University of California at Berkeley,
Berkeley CA 94720 USA }
\maketitle
\draft
\begin{abstract}
We study the single-vortex solution of the t-J model within resonating-valence-bond (RVB) mean-field theory. We
find two types of vortex cores, insulating and metallic, depending on the parameters of the model. The pairing
order parameter near both cores have  $d_{x^2 -y^2}+i\eta d_{xy}$ symmetry. For some range of $t/J$ the calculated
tunneling spectrum of the metallic vortex core agrees qualitatively with the STM tunneling data for BSCCO.
\end{abstract}
\narrowtext

Recent STM experiments\cite{fischer,pan} on vortex states in high $T_c$ cuprates have revealed many interesting
properties that are not adequately described on theoretical grounds.  Most of the existing literature relies on
conventional Bogoliubov-de Gennes (BdG) aproach with a $d$-wave order parameter\cite{kallin,wang,franz,yasui}. In
the mean time it is widely appreciated that the high temperature superconductors are doped Mott insulators, and
there is no a priori reason for the conventional BdG description to be applicable there.

In the present paper we report the first {\it unrestricted} (see below) mean-field theory study  for the vortex
states in the t-J model with Coulomb interaction. The main results are as follows. 1) Depending on parameters of
the model there exists {\it two} types of vortices, one with insulating core and the other with metallic core. At
a fixed $t/J$ and Coulomb interaction strength, the insulating core is favored by low doping while the metallic
core is favored by high doping. 2) Near the core of the vortex the pairing order parameter has $d_{x^2-y^2}+i\eta
d_{xy}$ symmetry (see later discussions). The value of $\eta$ tends to increase with doping. 3) The total
integrated single-electron spectral weight is proportional to the local concentration of holes. (In reality the
lost spectral weight will appear at energies above the charge gap which is taken to be infinite in the t-J model.)
4) The details of the tunneling spectrum inside a metallic vortex core, such as the existence of  zero-bias
peak\cite{wang,franz,yasui}, depends sensitively on the parameter choices for  $t/J$, Coulomb strength, and
doping. However the gross feature that the coherence peak tends to be suppressed in exchange for low-lying
spectral weight is observed in all cases. For the insulating core, the spectral weight is zero due to the Mott
constraint. For both cores the background $d$-wave behavior is recovered within a few lattice spacings from the
center of the core.

In standard notation, the Hamiltonian of the t-J model is given by
\be
H=-t\sum_{\ij} (c_{j\sigma}^\dag c_{i\sigma} \!+ h.c.)+J\sum_{\ij}(\Si\cdot\Sj \!-\frac{1}{4}\!n_i n_j).
\ee
In addition, due to the strong on-site repulsion the low-energy Hilbert space is constrained to have no more than
one electron per site.

So far this Hamiltonian has evaded exact solution in space dimensions greater than one. Limited exact
diagonalization results are often too small in system size to make a statement about complex experimental
situations. Meanwhile quite a lot is known about the various mean-field states of this
model\cite{affleck1,kotliar}. In particular the qualitative prediction of the phase diagram by the ``RVB''
mean-field theory\cite{kotliar} agrees with the experimental findings. Recently one of us showed that in the
superconducting state the gross prediction of the mean-field theory survives the low-energy gauge fluctuations
\cite{lee}.

In this paper we extend the mean-field treatment of Ref.\cite{kotliar} to study a single superconducting vortex.
Since the presence of a vortex breaks the translational symmetry, we allow all the mean-field order parameters to
be site/bond-dependent.   We call such calculation an ``unrestricted" mean-field theory. A similar approach has
been used recently to study the stripe phase of the t-J model\cite{vojta}. We believe that this mean-field theory
should be adequate to describe the relatively high-energy and short-distance physics of the vortex core.

Our starting point is the boson-fermion Lagrangian \widetext \ba 
L&&=\sum_i\{
\bar{b}_i(\partial_t\!-\!i\lambda_i \!-\!\mu)b_i+\fbar_{i\sigma}(\partial_t \!-\!i\lambda_i)f_{i\sigma}\}
 -t\sum_{\ij}(b_i\bar{b}_j\fbar_{i\sigma}f_{j\sigma}+h.c.) \!+\! {V_c\over 2}\sum_{i\ne j}\frac{1}{r_{ij}}\bbar_i b_i \bbar_j b_j \nonumber \\
&&+\frac{J}{4}\sum_{\ij} \{\Dbarij \Dij\!+\!\Kbarij \Kij \!- \!(\Dbarij \Pij + h.c.) \!-\!(\Kbarij \Hij +h.c.)\!
-\! n_i n_j\}. \label{act} \ea \narrowtext \noindent In the above $\Pij=\epsilon_{\sigma
\sigma'}f_{i\sigma}f_{j\sigma'}$, $\Hij=\fbar_{j\sigma}f_{i\sigma}$ (summed over spin), $\lambda_i$ is the
Lagrange multiplier that ensures the occupancy constraint, and $\Dij$ and $\Kij$ are Hubbard-Stratonovich fields.
 In the mean-field
approximation $b_i, \lambda_i, \Kij, \Dij$ are all treated as time-independent classical fields that minimize the
action. Note that we have also included the long-ranged Coulomb interaction in the model. Since we are treating
$b_i$'s as classical fields, the Coulomb interaction is in effect incorporated in the Hartree approximation.
Making use of the invariance of the action under local gauge transformation $b_j\rightarrow e^{i\phi_j}b_j$ and
$f_j\rightarrow e^{i\phi_j}f_j$, we will restrict $b_i$ to be real and positive. The values of $t/J$ and $V_c/J$
appropriate for the cuprate superconductors are not known exactly. In this paper we study a range of values for
such parameters and several doping concentrations.

The homogeneous mean-field ground state of the above action  is the $d$-wave superconducting phase first  worked
out by Kotliar and Liu\cite{kotliar}. In our notation, such ground state corresponds to $\lambda_i =\lambda$, and
$b_i =\sqrt{x}$ at every site ($x=$hole doping). To test our calculation we first study the homogeneous system
using the unrestricted mean-field theory. We find that for doping concentration $x$ that is neither too small nor
too large the solution is the uniform $d$-wave superconducting phase\cite{kotliar}. As a typical example,  the
electron density in the central $12\times 12$ region is shown for $t/J=1.25$, $V_c/J=1.25$ and $x=30/256$ in
figure 1(a). Due to the open boundary condition used, the electrons tend to accumulate at the edge (not shown),
leaving a somewhat higher hole density in the interior. We note that the electron density actually drops before it
increases. When the same unrestricted mean-field theory is applied to the insulating limit, $x=0$, we find the
plaquette-valence-bond state first discussed by Affleck and Marston\cite{affleck1}. The fact that we do not get
the N\'{e}el state  for zero doping is a well-known artifact of the RVB mean-field theory. At large enough doping
($x>x_c$) we find $\Dij=0$ and the solution describes a Fermi liquid.

We now turn to the case of a $hc/2e$ vortex in the superconducting state. In the following we shall only present
our $16\times 16$ results for $t/J=1.25$. We single out  $t/J=1.25$ in our study because in its neighborhood the
calculated tunneling spectrum agrees qualitatively with the experimental findings. For considerably larger $t/J$
the calculated tunneling spectrum deviates from the experimental measured one. For example for $t/J=3$ and
$x=30/256$ we find a zero-bias peak in the tunneling spectrum. It is clear that  we are more interested in results
that are independent of the parameter choice. The rest of the parameters we use are $0.125\le V_c/J\le 1.25$, and
$16/256\le x\le 38/256$.

The vorticity is imposed via the initial parameters \be K_{ij}=K_0,~~~ \Dij=\Dij^{0}e^{i\theta_{ij}},
b_i=\sqrt{x},~~~ \lambda_i=\mu_0 \label{eq:single_vortex} \ee where $K_0,\Dij^0,\mu_0$ are the bulk mean-field
parameters ($\Dij^0$ has $d_{x^2-y^2}$ symmetry), and $\theta_{ij}$ measures the angle made by the position vector
of the center of the $\ij$ bond and a fixed axis. We subsequently update the above parameters self-consistently.
Upon reaching self-consistency we find that $\Dij=\Dij^{'}e^{i\theta_{ij}}$, with $\Dij^{'}$ having
$d_{x^2-y^2}+i\eta d_{xy}$ symmetry\cite{note}.

One important feature of our mean-field solution is that it obeys \ba
&&\langle\bar{b}_ib_i\rangle+\langle\bar{f}_{i\sigma}f_{i\sigma}\rangle=1,~~~\forall i\nonumber \\
&&(K_{ij}\!+\!t\langle\bar{b}_jb_i\rangle)\langle\bar{f}_{i\sigma}f_{j\sigma}\rangle+
 t K_{ij}\langle\bar{b}_ib_j\rangle-c.c.=0,~~~\forall \ij.
\ea These equations imply that, on average, the total boson and fermion 3-currents do not fluctuate. This is a
direct consequence of the strong correlation physics inherent in the t-J model. In the field-theoretic
treatment\cite{lee}, a space-time local version of the above constraints arises as a consequence of integrating
out the gauge fluctuation.

A main conclusion of our work is the existence of {\it two types of vortex cores} - one with an insulating and the
other with a metallic core.  Small $V_c$ and low doping density favors the insulating core, and otherwise the
metallic core is favored.
\widetext
\begin{figure}[ht]
\hskip -0.5cm \centering \epsfxsize=15cm \epsfysize=4.5cm \epsfbox{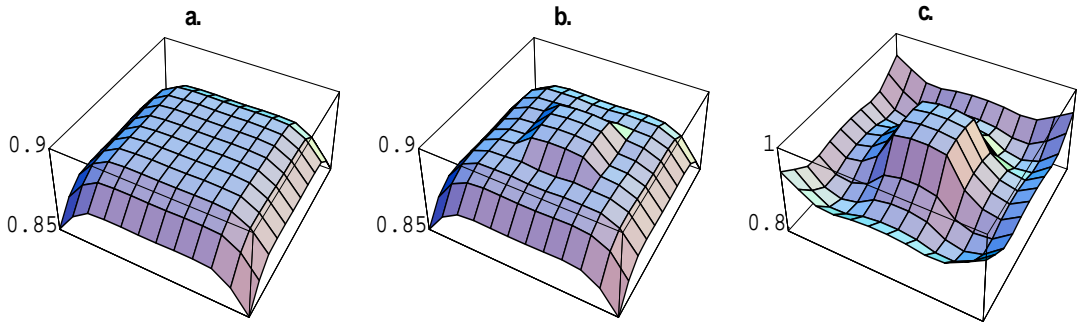} \caption{Electron density profile for
(a) no vortex ($x=30/256$, $V_{c}/J=1.25$), (b) vortex with metallic core ($x=30/256$, $V_{c}/J=1.25$), and (c)
vortex with insulating core ($x=16/256$, $V_{c}/J=0.125$). Only the central $12\!\times\! 12$ plaquettes of the
$16 \times 16$ lattice is shown.} \label{fig:den}
\end{figure}
\narrowtext
\noindent
 In Table \ref{table} we summarize our findings for a number of different parameter
choices. Based on this study we propose that at a fixed Coulomb interaction strength, {\it the vortex core changes
its nature from metallic to insulating as the system is progressively underdoped}.

{\bf{Vortex with Metallic Core:}} In figure 1(b) we show a typical example of a metallic-core vortex. We plot the
electron density profile near the center of a vortex for $V_c/J=1.25$ and $x=30/256$. First we note a slight
increase of electron density in the vortex core. In figure 2(b) we show the local density of states (DOS)
associated with a site on the central plaquette of the same vortex. Compared with the bulk DOS (black curve) there
seems to be more states at low energies in the vortex core (red curve). As we step away from the central plaquette
the two curves become indistinguishable beyond 4-5 lattice spacings. As mentioned earlier the self-consistent
order parameter $\Dij$ near the vortex core shows $d_{x^2-y^2}+i\eta d_{xy}$ symmetry\cite{yasui}. For $x=30/256$
and $38/256$, $\eta$ is about $30\%$ in the immediate vicinity of the core center. For $x=24/256$ and $x=16/256$,
we find $\eta\approx 20\%$ and $3\%$ respectively. Presumably a larger circulating current associated with higher
doping is responsible for this change in $\eta$. The appearance of the $id_{xy}$ component inside the vortex and
its consequence on the tunneling spectra were considered by several authors\cite{franz,laughlin,balatsky}.

Inside the metallic vortex core, the low-lying DOS depends sensitively on the doping level. This is illustrated in
figure 2 where we fix $V_c/J$ at 1.25 and vary the doping level among $x=16/256, 30/256$, and $38/256$. For
$x=16/256$, some of the spectral weights for $|E|\ge J$ is lost, while the lower energy portion is nearly
unchanged. The integrated DOS is clearly reduced by the vortex. For $x=30/256$, much of the lost spectral weight
under the peaks at $E=\pm J$ has re-emerged at lower energies. For $x=38/256$, the peaks at $E=\pm J$ are entirely
gone, and the DOS near $E=0$ is considerably higher. We believe that the low-lying DOS profile is determined by
the following two factors. 1)The presence of the $id_{xy}$ component opens up a gap and hence pushes the states
away from $E=0$. 2)The circulating current Doppler-shifts\cite{volovik} the quasiparticle energy levels and
increases the DOS at $E=0$. The dependence of the detailed shape of DOS on the choice of $t/J, V_c/J, x$ simply
reflects the variation of the above two factors with the parameters. \widetext
\begin{figure}[ht]
\centering \hskip -0.4cm \epsfxsize=15cm \epsfysize=4cm \epsfbox{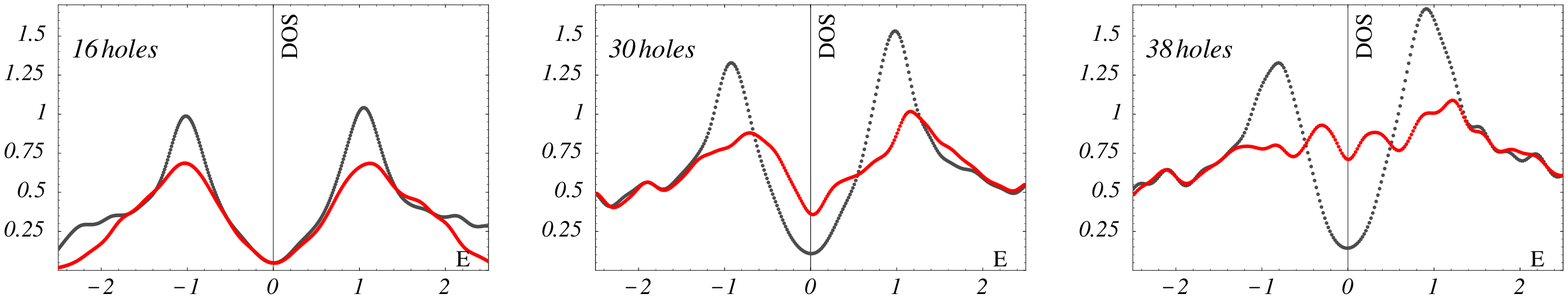} \caption{Density of states (red) at the
center of the metallic vortex core for $x=16/256,30/256$, and $38/256$ holes and $V_{c}/J=1.25$. Dark curves
represent the pure $d$-wave results for the same parameters.} \label{fig:dos}
\end{figure}
\narrowtext

When comparing the local DOS in the vortex core and far away, it is important to bear in mind that in the present
description the total integrated electron local DOS is proportional to $|b_i|^2$. Since the vortex core has a
higher electron density compared with the bulk, the corresponding $|b_i|^2$ is smaller. As a result it will appear
that some electron spectral weight has simply disappeared in the vortex core.  In reality, the lost spectral
weight should appear at energies greater than the Hubbard gap, which is treated as infinity in the present model.

\begin{figure}[t]
\hskip -0.4cm \epsfxsize=8cm \epsfysize=4cm \epsfbox{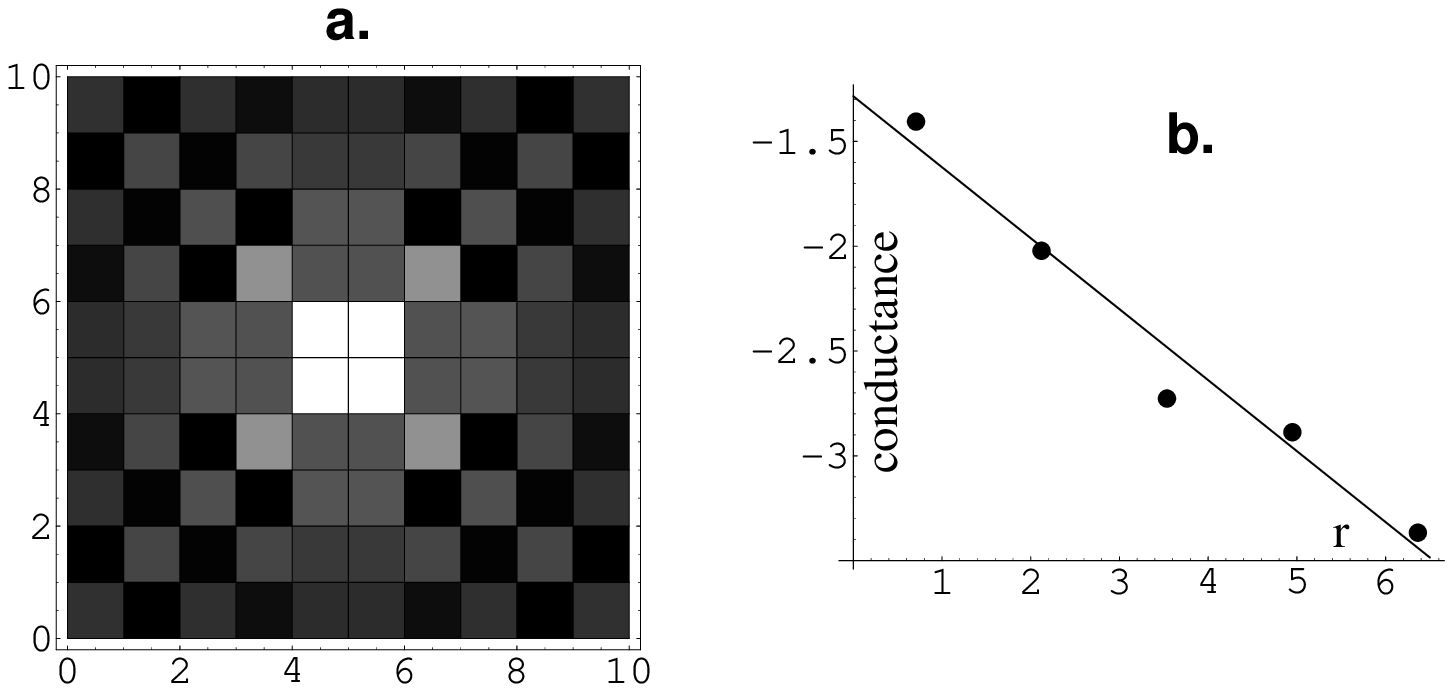} \caption{(a) Tunneling conductance difference
between vortex and non-vortex states. The central $10 \times 10$ lattice is shown where each site is represented
by a plaquette (dual lattice). Bright(dark) areas have a larger(smaller) conductance enhancement for the vortex.
(b) Semi-log plot of the conductance difference versus distance (lattice spacing=1) along the nodal direction. The
linear fit is made with a slope of $-0.34$.} \label{fig:cond}
\end{figure}

The extra low-energy DOS induced by the vortex diminishes with distance from the core. Experimentally Pan {\it et
al.} have measured the excess conductance at a fixed bias voltage in BSCCO and found that it decays
exponentially\cite{pan}. For the system with $x=30/256$ and $V_c/J=1.25$ (Fig.\ 2(b)), we have calculated the
extra DOS at a fixed energy, $E=0.3J$. Figure \ref{fig:cond}(a) shows such ``excess conductance" for the center
$10\!\times\! 10$ sites. We have also plotted the conductance along the diagonal (nodal) direction with  distance
in Fig.\ 3(b), where the straight line is a pure exponential behavior. The fall-off distance deduced from this
plot is approximately three lattice spacings.  The excess conductance shows considerable amount of angular
variation, with the maximum occurring along the nodal directions. Such anisotropy was not seen within the
experimental resolution\cite{pan}.

Clearly we do not intend to compare our results quantitatively with the measured ones. It is however significant that there is a range of (reasonable) parameter choice that
yields results in qualitative agreement
with the experimental findings. For this reason we believe that the model we study does capture the
essence of the vortex physics in the cuprates.
{\bf{Vortex with Insulating Core:}} In figure 1(c) we show a typical example of an insulating-core vortex, found
for $V_c/J=0.125$ and $x=16/256$. In this case the electron density in the central $4\times 4$ plaquette nearly
reached one. The coupling of the central insulating region and the surrounding is very weak, i.e. the links
connecting any of the core sites with the non-core sites have $\Dij\approx 0$, $\Kij\approx 0$. Since
$|b_i|^2\approx 0$ in the core there is negligible single-electron spectral weight there. The spin excitation
spectrum in the vortex core is almost the same as that of the plaquette-valence-bond state of the insulating
limit, i.e. a spin gap approximately equal to $2J$ separates the ground state singlet from the first excited
state. It is possible that this feature is an artifact of the inadequacy of the RVB mean-field theory to correctly
describe the N\'(e)el state. Outside the vortex core $|b_i|^2>0$ and the electron local DOS gradually recovers its
bulk value. As in the metallic case the integrated spectral weight suffers a factor of $|b_i|^2/x$ suppression
compared with the bulk. In the vicinity of the core we observe $d_{x^2-y^2}+i\eta d_{xy}$ order parameter. For
underdoped materials, the evolution of DOS in the above-mentioned manner as the tunneling tip moves away from the
center of the vortex will be an indication of the existence of insulating core. The possibility that for low
doping the vortices can have insulating cores has been discussed in several earlier works\cite{lee,so5,franz00}.
The existence of this type of vortex will be a clear manifestation of the proximity of the superconducting system
to the Mott insulator. We believe that the insulating vortex core and the appearance of charge stripes in the
underdoped cuprates have a common origin -- at a particular doping density the system is on the brim of phase
separation\cite{stripe,vojta}.

We conclude by noting that, since low-energy-integrated spectral weight is directly proportional to the local hole
density, {\it a careful measurement of its variation can, in principle, reveal the local charge distribution.}
This suggests an interesting possibility of determining the presence of a vortex or a charge stripe by direct
imaging in STM experiments.

We are grateful to G. Baskaran, Seamus Davis, Eric Hudson, Steve Kivelson, S.-H. Pan, Subir Sachdev, Matthias Vojta, and
Ziqiang Wang for valuable discussions, and to Marcel Franz for sending his manuscripts prior to publication. We
are particularly grateful to Ned Wingreen for numerous insightful remarks and questions. We also wish to thank NEC
research for the use of their computing facility. DHL is supported by NSF grant DMR 99-71503.

\begin{table}
\caption{Nature of vortex cores for various doping (horizontal entry) and Coulomb interaction strengths  (I=insulating, M=metallic).}
\begin{tabular}{cccc} $V_{c} /J$ &16/256 &30/256 &38/256  \\ \hline
1/4   &I  &I &M \\
3/4   &M   &M  &M \\
5/4   &M &M &M
\end{tabular}
\label{table}
\end{table}

\widetext


\begin{references}
\bibitem{note} The symmetry of $\Dij'$ is deduced by studying its
transformation property under $C_{4v}$ about the center of the vortex. According to this definition the result of
Ref.\cite{kallin} should also be interpreted as $d_{x^2-y^2}+i\eta d_{xy}$.
\end{references}
\end{document}